\documentclass{article}

\usepackage{arxiv}

\usepackage[utf8]{inputenc} 
\usepackage[T1]{fontenc}    
\usepackage[UKenglish]{babel}
\PassOptionsToPackage{hyphens}{url}
\usepackage{hyperref}       
\usepackage{url}            
\usepackage{booktabs}       
\usepackage{amsfonts}       
\usepackage{nicefrac}       
\usepackage{microtype}      
\usepackage{graphicx}
\usepackage[numbers,sort&compress]{natbib}
\usepackage{doi}
\usepackage{xcolor}

\definecolor{mygreen}{HTML}{2E7D32}

\title{Converting a superseded Raith electron beam pattern generator into an unattended Critical Dimension SEM}

\date{October 15, 2026}

\author{
	\href{https://orcid.org/0009-0000-4005-7907}{\includegraphics[scale=0.06]{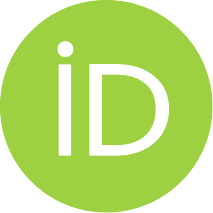}\hspace{1mm}Will~Gunn}\textsuperscript{1,}\thanks{Corresponding author: \texttt{w.gunn@uq.edu.au}}\quad 
    \href{https://orcid.org/0000-0002-4663-5111}{\includegraphics[scale=0.06]{orcid.pdf}\hspace{1mm}Han-Hao~Cheng}\textsuperscript{2} \quad
	\href{https://orcid.org/0000-0001-9476-5968}{\includegraphics[scale=0.06]{orcid.pdf}\hspace{1mm}Daniel~Peace}\textsuperscript{3} \quad
	\href{https://orcid.org/0000-0002-9053-9792}{\includegraphics[scale=0.06]{orcid.pdf}\hspace{1mm}Fernando~Gottardo}\textsuperscript{3} \\
	\textbf{\href{https://orcid.org/0009-0006-1828-1658}{\includegraphics[scale=0.06]{orcid.pdf}\hspace{1mm}Lachlan~Hanrahan}\textsuperscript{1} \quad
	\href{https://orcid.org/0000-0002-3049-6639}{\includegraphics[scale=0.06]{orcid.pdf}\hspace{1mm}Zachary~Degnan}\textsuperscript{1} \quad
	Keith~Moulding\textsuperscript{4} \quad
	Nit~Taksatorn\textsuperscript{5}} \\
	\textbf{\href{https://orcid.org/0000-0003-2203-3674}{\includegraphics[scale=0.06]{orcid.pdf}\hspace{1mm}Roger~Wepf}\textsuperscript{2} \quad
	\href{https://orcid.org/0000-0002-4659-3804}{\includegraphics[scale=0.06]{orcid.pdf}\hspace{1mm}Markus~Rambach}\textsuperscript{1} \quad
	\href{https://orcid.org/0000-0001-9639-5200}{\includegraphics[scale=0.06]{orcid.pdf}\hspace{1mm}Andrew~G.~White}\textsuperscript{1,3}} \\[2ex]
	\textsuperscript{1}School of Mathematics \& Physics, The University of Queensland, Brisbane, QLD, Australia \\
	\textsuperscript{2}Centre for Microscopy and Microanalysis, The University of Queensland, Brisbane, QLD, Australia \\
	\textsuperscript{3}National Advanced Photonics Facility, The University of Queensland, Brisbane, QLD, Australia \\
	\textsuperscript{4}Raith Asia Ltd, Hong Kong \\
	\textsuperscript{5}GenISys GmbH, Germany
}

\renewcommand{\headeright}{Preprint}
\renewcommand{\undertitle}{Preprint}
\renewcommand{\shorttitle}{Converting a superseded Raith E-beam pattern generator into a CD-SEM}

\hypersetup{
pdftitle={Converting a superseded Raith electron beam pattern generator into an unattended Critical Dimension SEM},
pdfsubject={physics.ins-det, physics.app-ph},
pdfauthor={Will Gunn, Han-Hao Cheng, Daniel Peace, Fernando Gottardo, Lachlan Hanrahan, Zachary Degnan, Keith Moulding, Nit Taksatorn, Roger Wepf, Markus Rambach, Andrew G. White},
pdfkeywords={Scanning electron microscopy, electron beam lithography, automated metrology, design-based metrology, Raith eLine Plus, silicon integrated photonics},
}

\begin{document}
\maketitle

\begin{abstract}
	Design-based metrology of academic electron beam lithography (EBL) is bottlenecked by low-throughput imaging. For integrated photonics, where nanometre-scale structural variation can significantly limit device performance, scanning electron microscopy (SEM) is the standard tool for critical dimension (CD) metrology, revealing the structure of devices that is not available from optical measurements. Manual SEM imaging on a wafer-scale is time-intensive, can feature operator-dependent distortions, and yields too few data to reliably close the nanofabrication calibration loop. The following details a methodology that utilises existing hardware in an academic cleanroom: a GDSII-to-SEM software workflow that converts a facility's superseded pattern generator into a CD-SEM for unattended image acquisition and efficient CD and line edge roughness (LER) extraction. This is shown for a Raith eLine~Plus, and is expected to be applicable to other pattern generators running Raith's \textsc{NanoSuite}. Presented are results demonstrating its three use cases within a facility: accelerated calibration of two photoresists' process windows; chip-scale metrology of integrated photonic devices to track CD variation; and mass dataset acquisition for convolutional neural network training to predict process distortion. This methodology enables foundry-style, metrology-driven process calibration in academic cleanrooms, generalisable beyond photonics to any EBL nanofabrication that requires nanometre-scale CD control.
\end{abstract}

\keywords{Scanning electron microscopy \and electron beam lithography \and automated metrology \and design-based metrology \and Raith eLine Plus \and silicon integrated photonics}


\section{Introduction}\label{sec:intro}

Metrology workflows in electron beam lithography (EBL) rely on scanning electron microscopy (SEM) for top-down critical dimension (CD) measurements of structures. In semiconductor manufacturing and EBL foundry services, design-based metrology (DBM) workflows marry a floorplan layout to the physical structures on a chip/wafer surface, enhancing quality control. Measurements in these settings are performed by critical dimension SEMs (CD-SEMs) that automatically image large datasets of structures on the wafer-scale \cite{Orji2018}, with DBM workflows setting up these imaging recipes directly from design data \cite{IRDS2023}. In shared nanofabrication facilities, where the cost of CD-SEMs is often not an option, metrology relies on manually operated SEMs. For facilities with access to commercial nanofabrication software suites, such as GenISys \textsc{ProSEM}, \textsc{Tracer} and \textsc{Beamer} \cite{GenISysProSEM,GenISysTRACER,GenISysBEAMER}, image analysis and process-aware design are no longer the bottleneck towards achieving best practice fabrication. In our experience, the largest obstacle for academic cleanrooms is the volume of time required to manually image samples, as manual SEM set-up is time-consuming and image quality depends strongly on user ability \cite{Cho2024}. While after-market upgrades exist that can add layout-driven automation to a manual SEM column  \cite{GenISysInSPEC}, the following shows how traditional bottlenecks towards achieving automated imaging workflows may be alleviated by utilising existing hardware in a facility. By combining a superseded Raith eLine~Plus electron beam (E-beam) pattern generator, image analysis software \textsc{ProSEM} \cite{GenISysProSEM}, process calibration software \textsc{Tracer} \cite{GenISysTRACER}, and custom automation scripts, we achieve an automated DBM workflow within an academic cleanroom.

This is possible since Raith E-beam pattern generators share many of the same components as CD-SEMs. Both consist of an SEM column with an in-lens/through-lens secondary electron detector \cite{RaithMultifunctionalEBL,RaithProImage}, the same through-lens detection used by CD-SEMs for high-resolution measurement \cite{Orji2018}. The accuracy of CD-SEM measurements is calibrated with reference samples \cite{Orji2018}. Similarly, a Raith pattern generator completes writefield alignment and correction with a laser-interferometer stage on a standard calibration chip (a `chessy') with checkerboard structures. These remain on an eLine~Plus sample holder and are used to align pattern generation writefields for EBL, and for calibrating SEM magnification, X and Y scaling and distortion. Combined with three-point rotation and working distance correction of the sample itself (Section~\ref{subsubsec:methods/image-acquisition}), every scanframe imaged through the pattern generator is spatially and focally corrected, and the eLine~Plus can be operated as a functionally equivalent CD-SEM. Semi-automated imaging is available on the eLine~Plus by defining scanning routines with the same position list structures used for pattern generation, but these lists typically require manual structuring and are the largest bottleneck towards automated metrology within our facility.

For small dataset acquisition, users report that manual structuring of imaging position lists on the eLine~Plus often takes longer than imaging the sample without automation (Section~\ref{subsec:methods/autosem-software}). Native automation on Raith EBL tools is designed for exposure \cite{RaithAutoRun}, and we find no prior work that generates Raith position lists for unattended imaging directly from a design floorplan. In response, we present \textsc{AutoSEM}, an open-source, website application-based program that converts a GDSII design floorplan into ready-to-scan position lists. This allows unattended image acquisition on a Raith eLine~Plus. With acquired images passed to GenISys \textsc{ProSEM}, \textsc{Tracer} and \textsc{Beamer} for process calibration and proximity effect correction (PEC) of future exposures, the result is an automated DBM workflow within our facility. By shifting the focus of users from manual imaging at the tool to defining scanframes at the design stage, we enable faster training, consistent image quality regardless of the user's SEM operation skills, and wider adoption of metrology workflows. The software and analysis methodology are detailed in Section~\ref{sec:methods}. Section~\ref{sec:results} demonstrates the workflow across three use cases: accelerated process calibration of two EBL photoresists; chip-scale metrology of silicon photonic devices planned at the design stage; and large-scale dataset acquisition enabling next-generation process calibration with convolutional neural networks. Limitations in the methodology, remaining open challenges, and visions for generalisations across nanofabrication are discussed in Section~\ref{sec:discussion}.

\section{Methods}\label{sec:methods}

The following describes the specifics of our automated DBM workflow, detailing software architecture and automated image acquisition (Section~\ref{subsec:methods/autosem-software}), and its integration with GenISys analysis software (Section~\ref{subsec:methods/genisys-software}). There are three categories of software used in this workflow. The first is software native to the Raith eLine~Plus, known as \textsc{NanoSuite}, with custom scripts for extended use. The second is an open-source, website application-based program developed to fully automate scanframe positioning, and the third is commercially available GenISys software \textsc{ProSEM}, \textsc{Tracer} and \textsc{Beamer}, used for image analysis, process calibration and patterning. Structures in this workflow are patterned with a Raith EBPG5150 at 100\,kV. All images are acquired with a Raith eLine~Plus in-lens secondary electron detector at 10\,kV acceleration voltage, 30\,$\mu$m aperture, $\sim$$300\,$pA probe current, $\sim$$10\,$mm working distance, and two-point averaging.

\subsection{Software architecture \& image acquisition workflow}\label{subsec:methods/autosem-software}

\subsubsection{Existing semi-automation in the Raith \textsc{NanoSuite}}
A Raith eLine~Plus is shipped with the native software \textsc{eLine~Plus}, part of Raith's \textsc{NanoSuite}, and \textsc{SmartSEM} developed by ZEISS. \textsc{SmartSEM} is a user interface (UI) for using the tool exclusively as an SEM, and \textsc{eLine~Plus} is for system control and pattern generation. SEM images on the system are usually acquired with \textsc{SmartSEM}, but if users image samples with \textsc{eLine~Plus}, preset scanframes that have fixed magnification, writefield size and imaging conditions can be defined in a scan manager. This raises the consistency of images across a sample. Preset scanframe images also benefit from the same writefield alignment and three-point rotation/working distance correction used during pattern generation (Section~\ref{subsubsec:methods/image-acquisition}). On an alignment-corrected sample with well-calibrated writefields, such imaging can reliably achieve sub-10\,nm resolution. For the three use cases (Sections~\ref{subsec:results/usecase1}, \ref{subsec:results/usecase2}, \ref{subsec:results/usecase3}), all images are acquired with 2\,nm/px resolution.

Moreover, the \textsc{eLine~Plus} software allows users to pass JavaScript (.js) commands that control stage movements and the scheduling of scanframe imaging. This scheduling occurs in a position list file (.pls) structure, the same structure used to schedule pattern generation, where the positions of exposures/images are associated with local coordinates on an aligned sample. In coordination with Raith, automation scripts are written, and preliminary attempts to integrate these scripts into position list structures give a semi-automated imaging protocol. This is considered semi-automated, since correctly structuring position list files is a non-trivial task for facility users. Manually assigning scripts to the correct scanning areas on a sample often takes longer than imaging the sample without automation. To address this, we develop a portable, website application-based program that automates position list structuring for fully automated, unattended imaging.

\subsubsection{Custom UI development to automate position list structure}

\begin{figure}[htbp]
    \centering
    \includegraphics[width=0.6\linewidth]{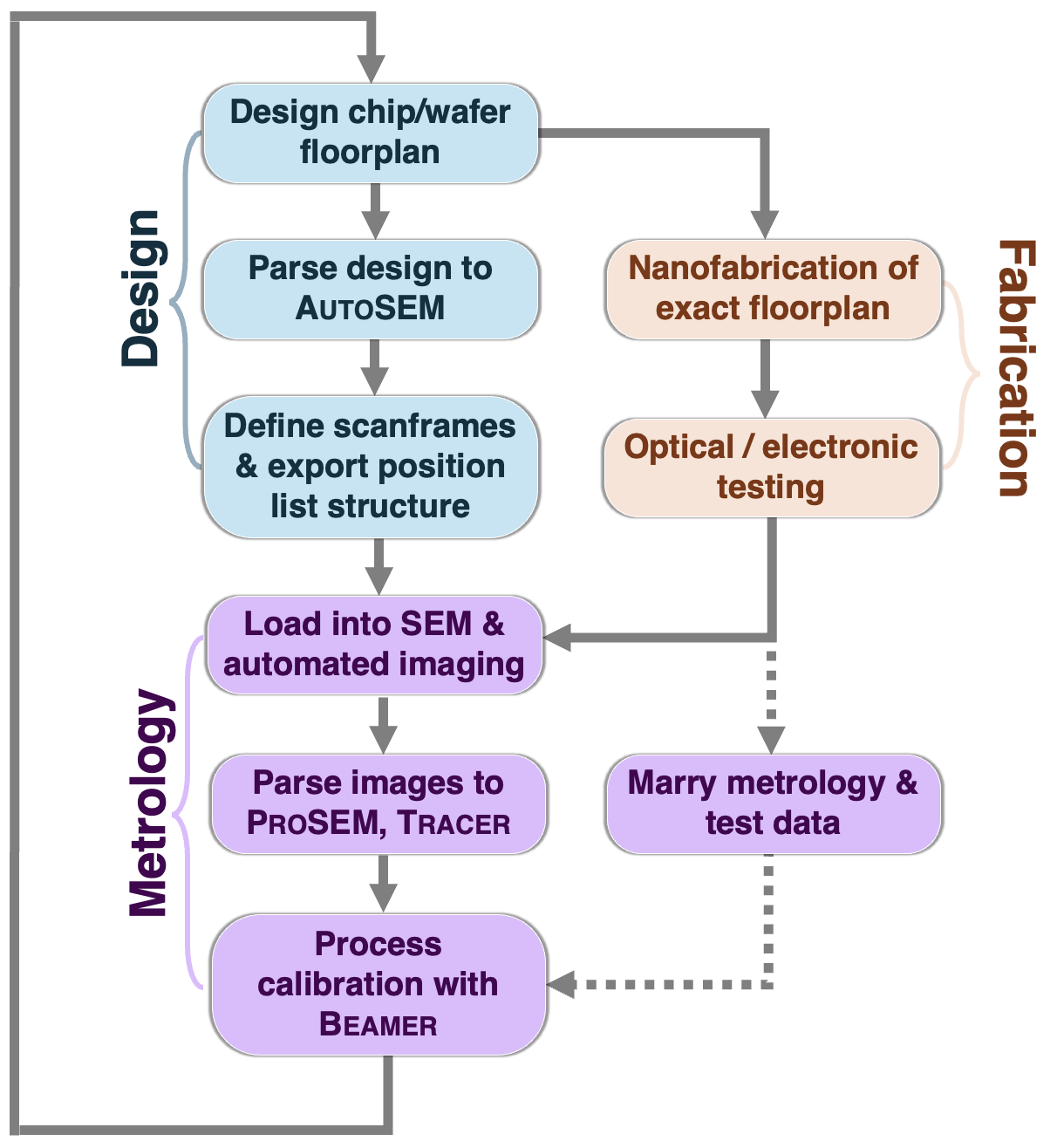}
    \caption{Overview of the design-based metrology workflow for a fabrication feedback loop. A 1-to-1 correspondence between design floorplan and fabricated floorplan enables layout-driven metrology. Concurrent to fabrication and testing, design floorplans are parsed into \textsc{AutoSEM} to automatically structure position lists of scanframes. After fabrication and chip/wafer loading into a Raith eLine~Plus or other Raith pattern generator, the GDSII design file and position list structures automate imaging. Datasets acquired are parsed to GenISys \textsc{ProSEM}, \textsc{Tracer} \& \textsc{Beamer} for process calibration to enforce future layout designs, closing the fabrication feedback loop.}
    \label{fig:autosem-workflow}
\end{figure}

The website application \textsc{AutoSEM} (open-source\footnote{Available at \url{https://github.com/wafgunn/AutoSEM}}) enables user-friendly, high-throughput imaging by ingesting users' GDSII design files and automating the creation of position lists. The \textsc{AutoSEM} codebase is packaged into a single HTML file for portability, and is designed as a UI split into three steps: a GDSII design file, a scanframe assigner and organiser, and an instruction exporter from which ready-to-scan position lists can be downloaded. Figure \ref{fig:autosem-workflow} visualises where \textsc{AutoSEM} exists within the fabrication feedback loop.

The GDSII viewer allows users to drag-and-drop their design file that contains a complete floorplan of their nanofabrication, navigate between cells, hide/show layers, and perform basic layout viewer commands to preview the design before organising metrology. A scanframe assigner allows users to place scanframes onto their floorplan from a list that matches preset scanframes stored in the \textsc{eLine~Plus} scan manager. The application restricts users to the preset scanframes to ensure well-aligned writefields during imaging for all users in a facility, which keeps imaging runs on the communal tool reproducible. Image quality depends on the user's skill when correcting for stigmation \cite{Cho2024} and writefield misalignment. Since we believe sub-10\,nm critical dimension resolution is needed for accurate imaging of fine-featured photonic devices, inconsistency between users reduces the quality of an image set. Secondly, restricting users to preset scanframes allows for standardisation of the workflow where scanframes of set dimensions can be defined during the design process of the original GDSII floorplan. Users are given a lookup table of layers/datatypes corresponding to the set scanframe dimensions (layer 500, datatypes 0--9). Box objects on these specific layers/datatypes can be automatically imported by \textsc{AutoSEM}. If users have a GDSII design that already contains bounding boxes at positions they wish to image, they can manually import these boxes and associate the location with any of the preset scanframes. A manipulation menu is used to place selections of scanframes into named groups, allowing users to determine imaging order and pass identifying metadata in the group name.

After placing all desired scanframes, users can export instructions with key imaging parameters. These include the pixel number, defining the image resolution, as well as any line or frame averaging. Included is the ability to set a waiting period between scans, reducing any image drift error. The user then exports a fully formatted position list containing all scanframes and parameters, ready to be imported into the \textsc{eLine~Plus} software. All scanframes within this position list, regardless of the imaging extent, are defined by a central coordinate around which the pattern generator scans. The coordinate system for the scanframes is automatically determined by the maximum extent of the user's GDSII design floorplan, and the alignment marker coordinates are expressed in this same frame. This enables users to create position lists with scanning instructions for full-chip or full-wafer imaging, so long as the user implements three-point rotation/working distance alignment on the pattern generator (Section~\ref{subsubsec:methods/image-acquisition}). Other exports in this step include a text file containing scan number, central coordinate ($x,y$), scanframe size, pixel step size ($\mu$m), and the text string of the scanframe's group name that can include additional notes or metadata. Additionally, the user is able to export a final GDSII file that includes both the original design floorplan and the scanframe box objects.

\subsubsection{Image acquisition setup on the Raith eLine~Plus pattern generator}\label{subsubsec:methods/image-acquisition}
\begin{figure}[htbp]
    \centering
    \includegraphics[width=0.6\linewidth]{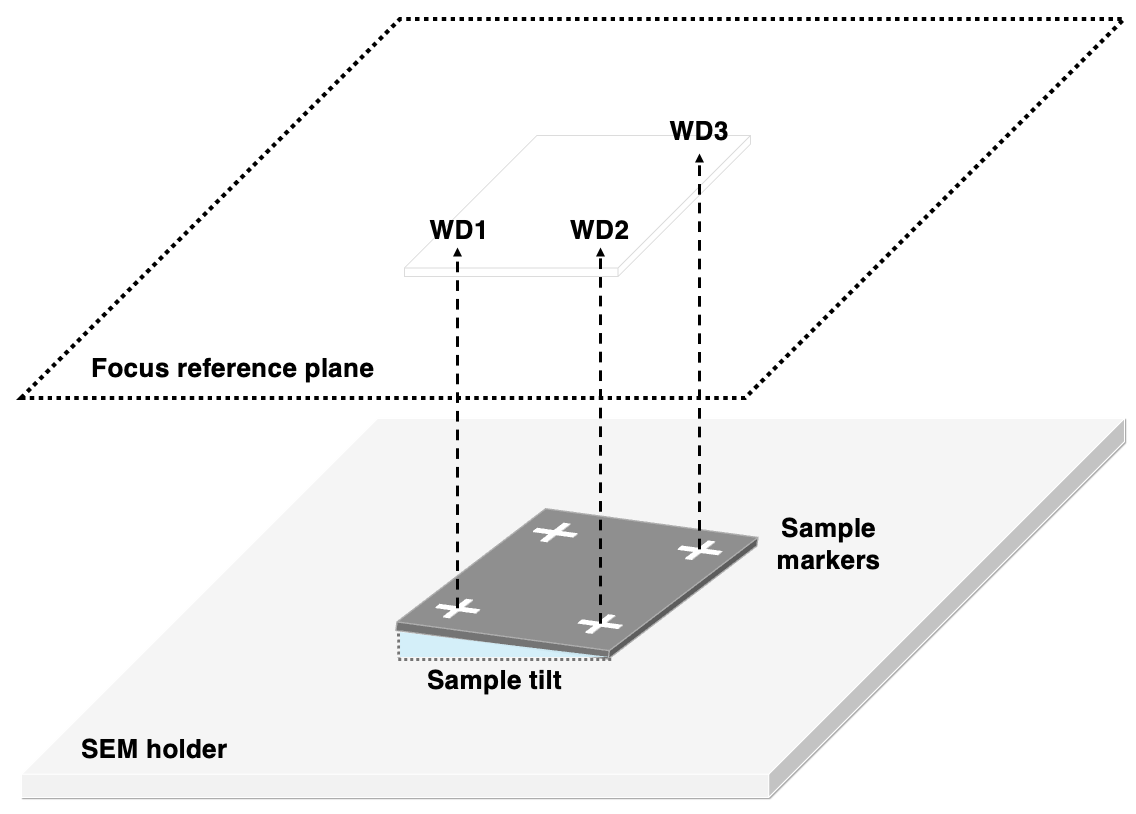}
    \caption{Positional focus correction. During three-point alignment, users ensure sharp focus at three non-collinear sample markers on a tilted sample, recording working distances WD1, WD2 and WD3, from which the software fits a plane and interpolates a working distance correction for every subsequent scanframe (Section~\ref{subsubsec:methods/image-acquisition}).}
    \label{fig:wd-correction}
\end{figure}

Before running the image acquisition on the pattern generator, a three-point alignment procedure is required to define a new local coordinate system and apply spatial/focal corrections. The \textsc{eLine~Plus} alignment takes three non-collinear fiducial/alignment markers---usually cross or L-shaped patterns---on the surface of the sample, and pairs these points with the corresponding coordinates in the user's GDSII design file. This corrects rotation, translation and axis scaling errors to the accuracy of the marker positions. To correct for any sample tilt, if users ensure sharp focus at each of the markers during alignment, the software calculates the slanted focal plane of the sample surface and interpolates working distance corrections for all future scans (Figure~\ref{fig:wd-correction}). By the end of the alignment procedure, all scanframe locations match true locations on the surface of a sample, and all images at these locations are spatially/focally corrected. Following alignment, the position list can be executed and all images can be reliably acquired unattended by the pattern generator. The text file exported from \textsc{AutoSEM} can be used as a lookup table for the image set, and a simple renaming script can include all necessary information from the group name text string into the file name.

\subsection{Batch analysis workflow for key fabrication metric extraction}\label{subsec:methods/genisys-software}

Following automated acquisition and renaming, large datasets of images are passed to GenISys software \textsc{ProSEM} for batch analysis \cite{GenISysProSEM}. Within \textsc{ProSEM}, a user can define regions of interest (ROIs) they wish to analyse, the types of expected features within those regions, as well as the pixel/nanometre length scale. From 8-bit greyscale values of an SEM image, the software can fit and analyse structure. A common example where batch analysis is used, to be discussed in further detail in Section~\ref{subsec:results/usecase1}, is dose-dependent isofocal analysis. When the image dataset is imported into \textsc{ProSEM} and the `Lines \& Spaces' structure preset is chosen, the software examines a single horizontal pixel slice and fits sigmoid functions to each edge of a line pattern. The top asymptotes of each sigmoid function are fitted to the brightest values within the pixel row and are associated with line edges, due to the bright edge bloom of SEM linescans \cite{SalvatPujol2019}. The base asymptotes of each function are fitted to the background grey level away from the edge. The inflections of each sigmoid are taken to be the central points of each line edge within a single row of pixels. \textsc{ProSEM} rasters each pixel row of the ROI and fits a straight line average through all inflection points to approximate line edges and an average linewidth. Comparing this width to the original design allows for dose-dependent CD variation to be estimated. Further, three times the standard deviation of the fitted inflection points away from each straight line average edge yields an estimate of LER ($3\sigma$). Fitting many ROIs over arrays of lines and spaces of varying densities enables the extraction of key fabrication metrics.

Measurements found in \textsc{ProSEM} are passed to GenISys \textsc{Tracer}, which fits the measured CDs to return the process bias to be applied to future device designs, the effective process blur (blur$_\mathrm{eff}$) and the base dose for PEC \cite{GenISysTRACER}. The isofocal dose is used as the base dose \cite{Lopez2017,Lopez2018}. These values are fed forward to PEC algorithms in GenISys \textsc{Beamer} \cite{GenISysBEAMER} to correct future exposures and close the fabrication feedback loop.


\section{Results}\label{sec:results}

The results from our automated DBM workflow are presented in three use cases, each serving a distinct task for EBL users within the nanofabrication facility. Use Case I (Section~\ref{subsec:results/usecase1}) details accelerated process calibration of two different EBL resists, including isofocal analysis, used to refine resist recipes. Use Case II (Section~\ref{subsec:results/usecase2}) demonstrates chip-scale imaging of silicon photonic devices, enabling metrology that tracks sub-10\,nm structural variation. Use Case III (Section~\ref{subsec:results/usecase3}) is a showcase of machine learning-based prediction and correction of fabrication variation, an emerging approach in silicon photonic design \cite{Gostimirovic2022,Gostimirovic2023}, now generalisable to individual steps within a fabrication recipe due to large-scale, unattended dataset acquisition. What ties these cases together is the significant acceleration and repeatability afforded by our automated DBM workflow compared to user-driven, manual processes.


\subsection{Use Case I: Accelerated EBL process calibration via systematic parameter sweeps}\label{subsec:results/usecase1}

Writing accurate patterns into EBL photoresists requires the creation of a process recipe, and writing consistent masks at the sub-10\,nm scale requires ongoing calibration of that recipe. Calibration methods track the individual steps and parameters that incur intolerable pattern distortions, arising from substrate cleanliness, photoresist film thickness, pre/post-exposure baking conditions, development, and etching conditions. The convolution of these distortions, as well as distortions resulting from patterning parameters (forward scattering, beam current and aperture, and beam step size) and tool artefacts (beam focus), is treated as an effective process blur (blur$_\mathrm{eff}$) \cite{Lopez2017,Lopez2018}. Best practice calibration tracks CD variation as a function of dose and pattern density, known as isofocal analysis \cite{Lopez2017,Lopez2018}. Isofocal analysis is conducted by patterning lines and spaces test structures that vary in fill-factor. To reduce backscattering between regions, each region is separated from its neighbours by more than $3\beta$, where $\beta$ is the lateral range of backscattered electrons for the photoresist/substrate combination. These patterns range from a single line structure ($\sim0\%$ density) up to an entirely filled region with a single space structure ($\sim100\%$ density), and include intermediate lines and spaces fill-factors (25\%, 50\%, 75\%). Imaging the linewidths across a dose range yields CD-versus-dose curves (exposure latitudes) for each pattern density \cite{Lopez2018}.

\begin{figure}[htbp]
    \centering
    \includegraphics[width=1\linewidth]{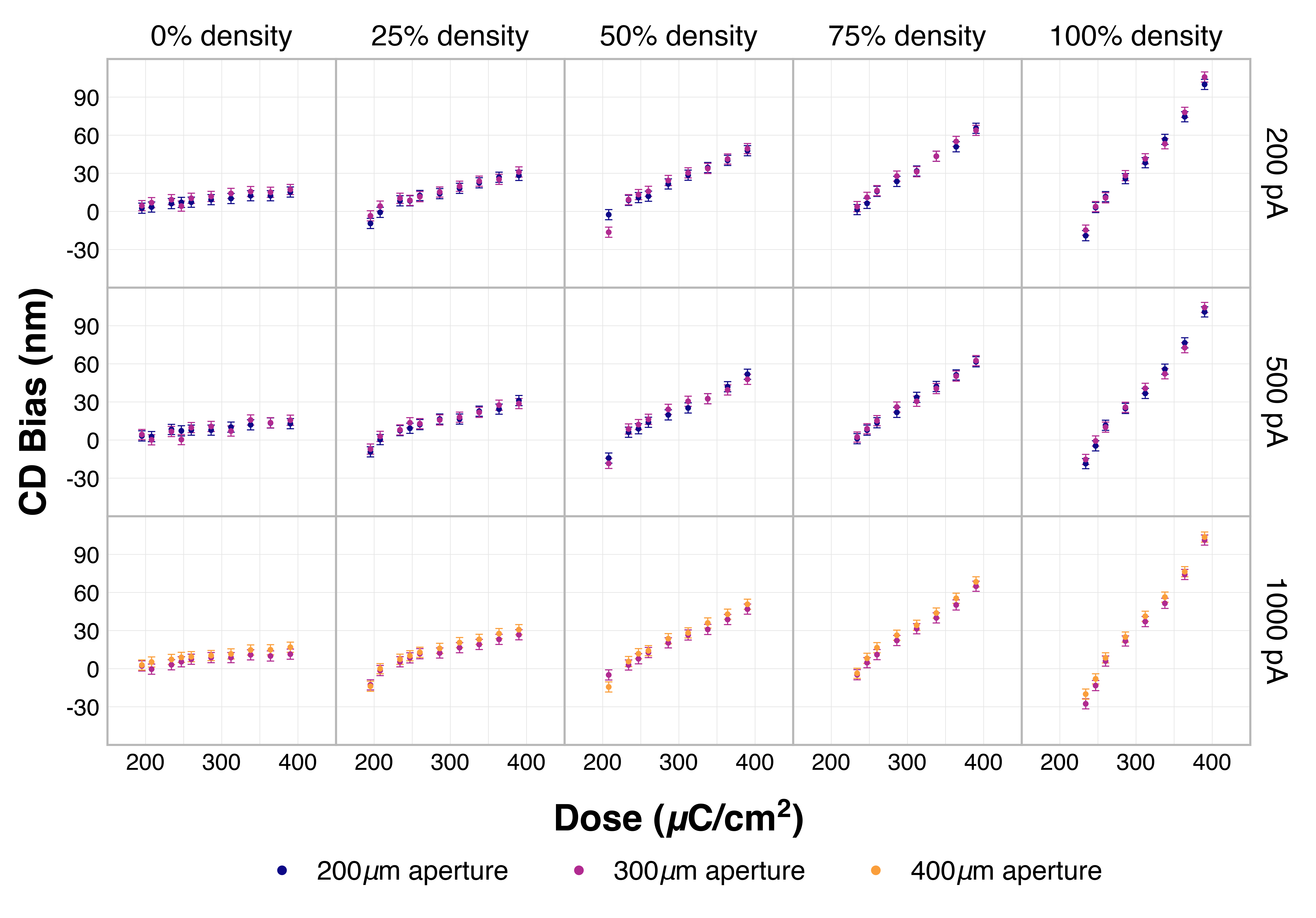}
    \caption{Exposure latitudes of \mbox{AR-P~6200.09} on SOI, showing CD bias as a function of dose for lines and spaces structures of five different densities (one density per column, from $\sim0\%$ to $\sim100\%$). Each row is a different beam current (200\,pA, 500\,pA and 1000\,pA). The 200\,$\mu$m (navy) and 300\,$\mu$m (magenta) apertures are compared at 200\,pA and 500\,pA, and the 300\,$\mu$m (magenta) and 400\,$\mu$m (orange) apertures are compared at 1000\,pA. Error bars are $\pm4$\,nm, a conservative estimate based on the 2\,nm/px pixel scale.}
    \label{fig:aperture-blur}
\end{figure}

\subsubsection{Isofocal calibration of \mbox{AR-P~6200.09} and \mbox{ma-N~2403}}\label{subsubsec:isofocal-calibration}

In this section, we detail how our automated DBM workflow enables acceleration and extension of isofocal analysis. Results below cover the automated calibration of two EBL photoresists on silicon-on-insulator (SOI): \mbox{AR-P~6200.09} by Allresist GmbH \cite{Allresist2026}, and \mbox{ma-N~2403} by micro resist technology GmbH \cite{MicroResist2026}.

For each resist, the dose-to-clear/dose-to-gel is determined beforehand from contrast curve analysis. An isofocal study is then completed over the resulting dose range, following the method outlined in Lopez~et~al.~\cite{Lopez2018}. Within a $100\times100\,\mu$m$^2$ exposure area, lines and spaces structures of five different densities ($\sim0\%$, 25\%, 50\%, 75\%, $\sim100\%$) are patterned across the fixed dose range, with a variety of different aperture and beam current configurations. All of these configurations are patterned on the same chip, reducing run-to-run variability of development and etching conditions. Each exposure area on chip is separated by a minimum of $100\,\mu$m distance ($\sim3.3\beta$) to reduce backscattering from neighbouring exposures. Following development, etching and resist removal, the samples are three-point aligned within the Raith eLine~Plus for image acquisition (Section~\ref{subsubsec:methods/image-acquisition}). From the GDSII design of the density structures, \textsc{AutoSEM} creates a chip-scale position list containing all patterning conditions, which is passed to the eLine~Plus. With this position list, 400 $5\times5\,\mu$m$^2$ images at 2\,nm/px resolution are acquired unattended across $\sim5.5$ hours. Using the lookup table generated by \textsc{AutoSEM}, the images are renamed to contain metadata related to beam conditions, dose and density, and parsed to \textsc{ProSEM}. CD and LER measurements are extracted for all images and passed to GenISys \textsc{Tracer} to fit estimates for process blur, process bias and isofocal dose.

\begin{figure}[htbp]
    \centering
    \includegraphics[width=0.6\linewidth]{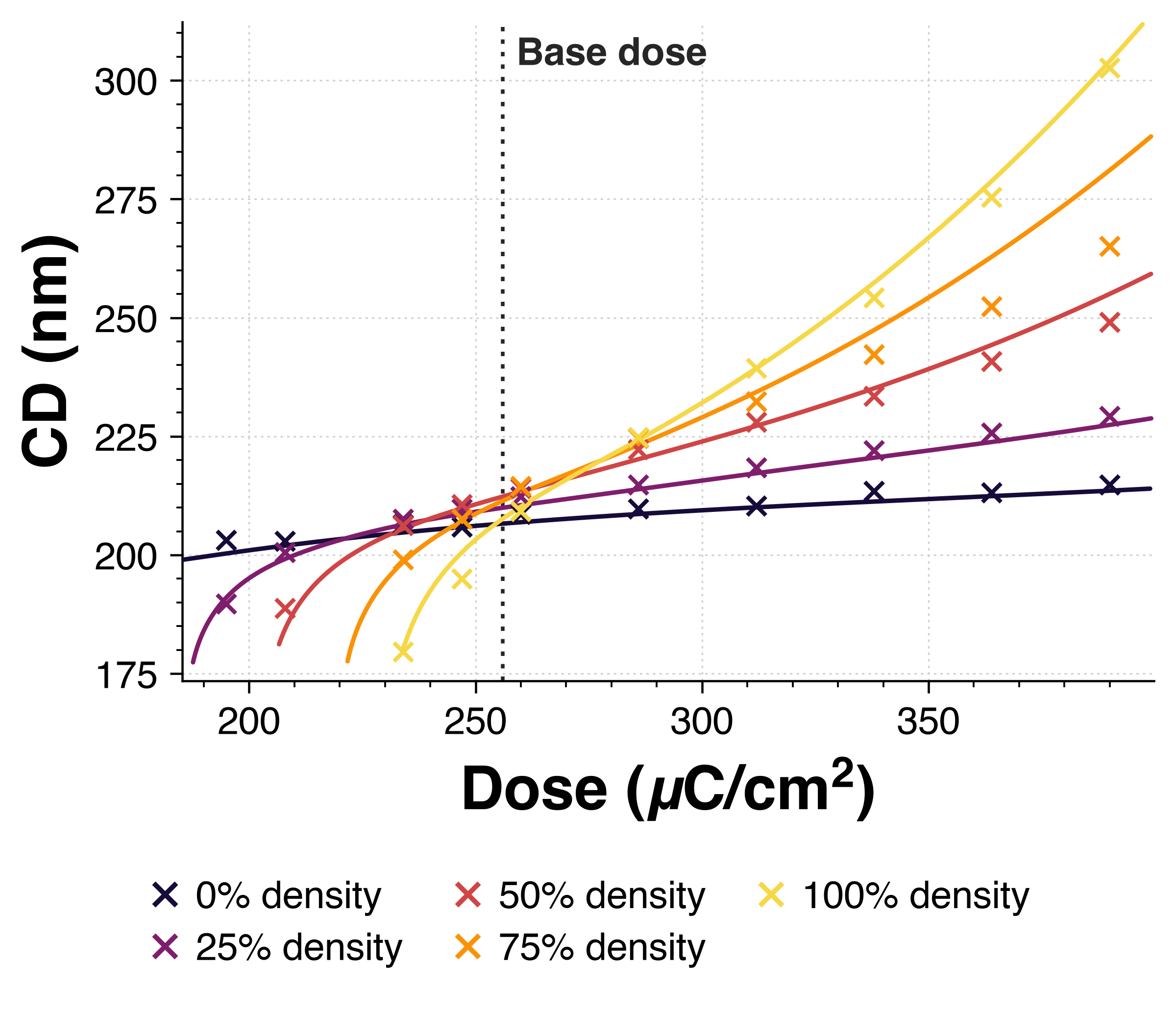}
    \caption{\textsc{Tracer} curve fitting for \mbox{AR-P~6200.09}, 500\,pA beam current, 300\,$\mu$m aperture. Crosses are CD measurements from \textsc{ProSEM} and lines are \textsc{Tracer} fits for the five different densities. The vertical line marks the fitted isofocal dose of $\sim$256$\,\mu$C/cm$^2$. \textsc{Tracer} also estimates a process blur of $\sim$$20$\,nm and a process bias of $\sim$$6$\,nm.}
    \label{fig:tracer-blur}
\end{figure}

Exposure latitudes for both \mbox{AR-P~6200.09} and \mbox{ma-N~2403} reveal dose-dependent CD variations. For brevity, the results of \mbox{AR-P~6200.09} are shown in Figure~\ref{fig:aperture-blur}. The latitudes demonstrate that CD bias varies as a function of dose for the five different pattern densities. For each density, a range of beam currents and apertures are considered that cover the typical conditions used for high resolution patterning. Error is conservatively estimated to be $\pm4$\,nm, accounting for the SEM pixel scale of 2\,nm/px. At 100\,kV and beam currents up to 1\,nA, the theoretical beam size can change by a factor of $\sim$1.5--2.5 between each pair of apertures shown in Figure~\ref{fig:aperture-blur} (200/300\,$\mu$m at 200 and 500\,pA, and 300/400\,$\mu$m at 1000\,pA). The invariance in trend between the 200\,$\mu$m, 300\,$\mu$m and 400\,$\mu$m apertures suggests that blur$_\mathrm{eff}$ is dominated by process contributions rather than by the beam spot size. This is supported by \textsc{Tracer} curve fitting (Figure~\ref{fig:tracer-blur}), finding an estimated process blur of $\sim$$20$\,nm, compared against beam sizes $\sim$$2.5$\,nm, $\sim$$4$\,nm and $\sim$$10$\,nm for each respective aperture at beam currents up to 1\,nA.\footnote{Beam spot sizes determined experimentally by our nanofabrication facility.} Alongside process blur, \textsc{Tracer} estimates an isofocal dose of $\sim256\,\mu$C/cm$^2$ and an incurred process bias of $\sim6$\,nm. Due to the invariance between beam current and aperture, this presents a robust process window across these patterning conditions centred around the isofocal dose. The same workflow applied to \mbox{ma-N~2403} finds a recommended isofocal dose of $\sim$$330\,\mu$C/cm$^2$, process blur of $\sim$$32$\,nm and process bias of 2\,nm, independent of the same beam current and aperture patterning conditions.

With this workflow, a wide variety of patterning conditions can be written across a single chip and imaged on a Raith pattern generator, with the associated position list generated automatically. The images are fed forward into \textsc{ProSEM}, and the results are used to enforce design correction for future fabrication with \textsc{Tracer} and \textsc{Beamer}. We estimate that this reduces the duration of the workflow from three days down to less than a single day. This is a result of considerably accelerated and unattended imaging, allowing the user to focus on the analysis and feed-forward corrections to future design.


\subsection{Use Case II: Chip-scale structural variability mapping for quality assurance \& quality control}\label{subsec:results/usecase2}

The next use case describes how automated position list structuring with \textsc{AutoSEM} enables metrology to be incorporated into the nanofabrication process from the design phase, for quality assurance and quality control of on-chip devices. The workflow allows more reliable tracking of the structural variability of inverse design photonic devices and photonic microring resonators. In both instances, nanometre-scale dimensional change impacts performance, shifting the resonance wavelength of filters and resonators \cite{Bogaerts2018}. The tolerance of hole-based inverse design photonic crossings has been analysed in simulation for $\pm10$\,nm changes in hole diameter \cite{Ma2024}.

\begin{figure}[htbp]
    \centering
    \includegraphics[width=0.7\linewidth]{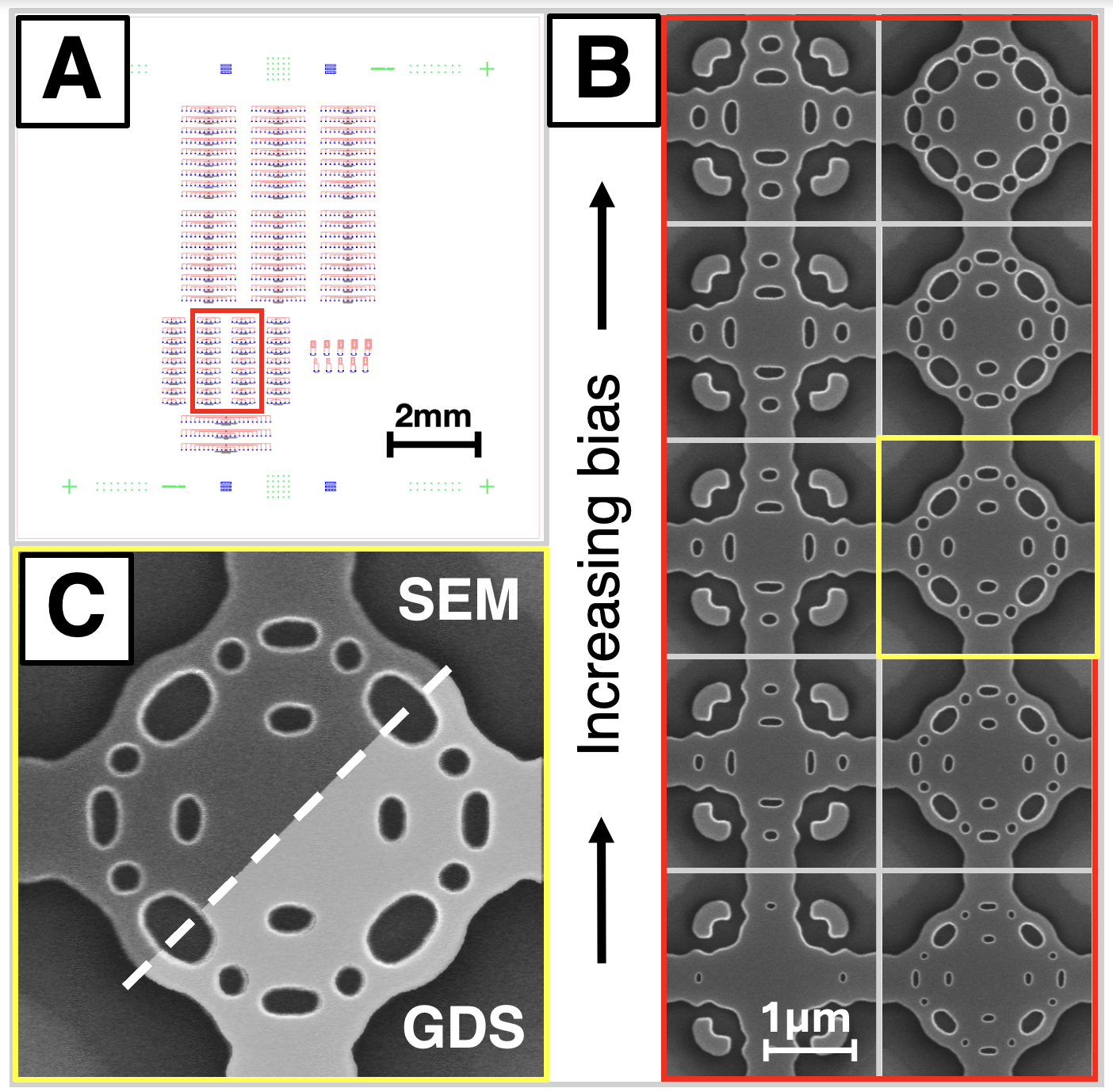}
    \caption{(A) GDSII floorplan of a silicon photonics chip featuring 103 individual components and 1001 grating coupler structures (scale bar 2\,mm). The red box marks the inverse design photonic crossings imaged in (B). (B) The 10 photonic crossings, covering two different device designs (left and right columns) at five different values of interior hole bias (increasing bottom to top). Each is imaged in a $5\times5\,\mu$m$^2$ scanframe at 2\,nm/px resolution and is shown cropped (scale bar 1\,$\mu$m). The yellow box marks the device shown in (C). (C) The same crossing, split along the dashed line into the SEM image (upper left) and GDSII design (lower right) to compare the critical dimension variation between design and device.}
    \label{fig:invdes-threeplot}
\end{figure}

\begin{figure}[htbp]
    \centering
    \includegraphics[width=0.9\linewidth]{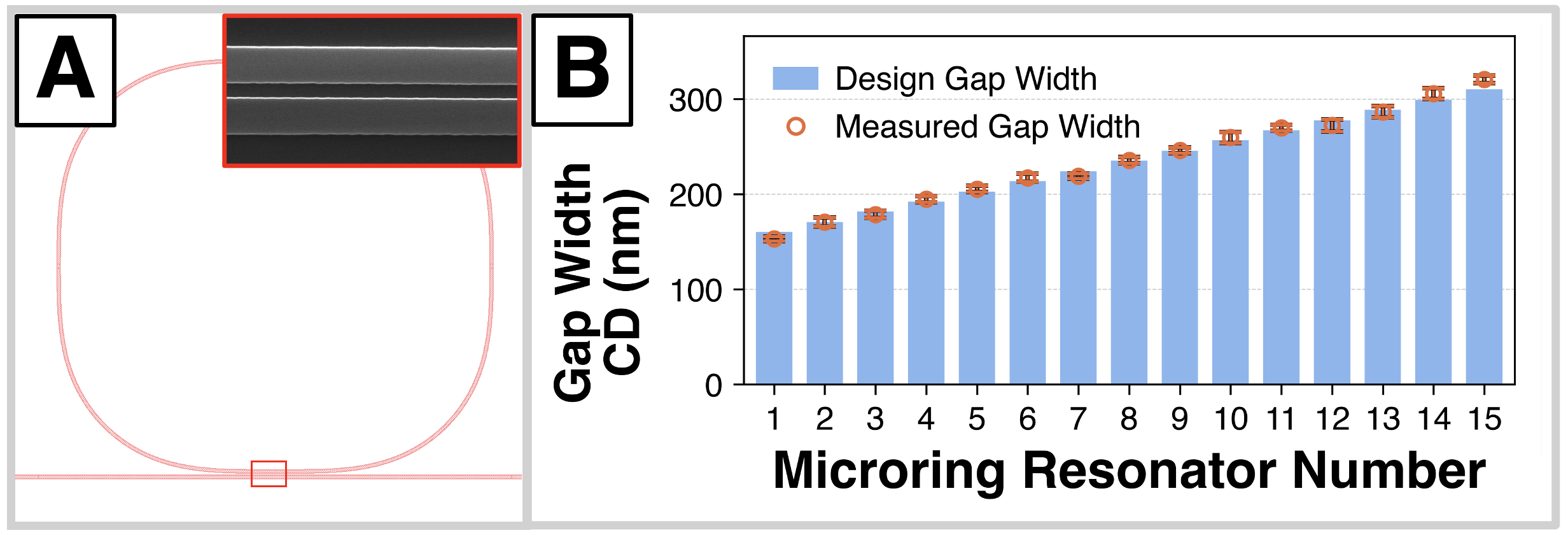}
    \caption{(A) GDSII design of a microring resonator (MRR) and bus waveguide. The red box marks the coupling region, and the inset shows an SEM image of the coupling gap, imaged during automated metrology. (B) Design (blue bars) coupling gap widths for the 15 different coupling gap values (MRR1--MRR15), and the measured widths (orange circles) extracted by \textsc{ProSEM}. Error bars are the standard deviation of the measured gap width from \textsc{ProSEM}.}
    \label{fig:gap-width}
\end{figure}

An example floorplan of a photonic chip is visible in Figure~\ref{fig:invdes-threeplot}A, featuring 103 individual components and 1001 grating coupler structures. Metrology of all 1104 on-chip structures can be organised during the GDSII design stage. When optical testing finishes, \textsc{AutoSEM} generates a position list ready for image acquisition. Highlighted in red in Figure~\ref{fig:invdes-threeplot}A, and shown in Figure~\ref{fig:invdes-threeplot}B, is a sample of 10 photonic crossings that cover two different device designs at five different values of interior hole bias. They are intended to test performance sensitivity of inverse design photonic crossings. Edge detection algorithms that fit sigmoid functions between the bright edge signal and the background grey level (Section~\ref{subsec:methods/genisys-software}) can be utilised to generate binary masks of devices and compare the critical dimension variation between design and device (one crossing is shown as a split SEM/GDSII view in Figure~\ref{fig:invdes-threeplot}C). This is a particular requirement for the interior features of inverse design devices, where nanometre-scale structural variation can affect transmission and cross-talk \cite{Gostimirovic2022,Gostimirovic2023,Ma2024}.

In Figure~\ref{fig:gap-width}A, the design for a microring resonator (MRR) is shown, and an inset shows an image of the MRR's coupling gap. For the MRRs in this study, the coupling gap is swept across 15 different values (MRR1--MRR15) to test performance. The user images each coupling gap by passing the GDSII design floorplan through \textsc{AutoSEM}, and the critical dimension and line edge roughness variation of the gaps are extracted by \textsc{ProSEM}. The design and measured widths of each coupling gap can be seen in Figure~\ref{fig:gap-width}B.

Our automated DBM workflow, with preset scanning conditions that improve reliability image-to-image, accelerates the search for fabrication-tolerant inverse design structures and fine-tuning of sub-10\,nm scale structures. Moreover, the ability to image all on-chip structures raises the standard of metrology thought to be tenable in academic nanofabrication facilities, allowing any trends in optical performance to be directly correlated to variation during fabrication processes. We believe this level of metrology automation can be easily extended from the chip-scale to the wafer-scale, with acquisition times no longer the rate-limiting factor of metrology.


\subsection{Use Case III: Large-scale dataset acquisition for convolutional neural network-based process prediction \& correction}\label{subsec:results/usecase3}

We next demonstrate a powerful use case for the workflow that widens the scope of process calibration in academic nanofabrication facilities. This involves training convolutional neural networks on SEM images of fabricated structures, based on the structural variability prediction and correction studies by McGill University and the National Research Council Canada \cite{Gostimirovic2022,Gostimirovic2023}. While their models were trained on datasets fabricated and imaged by a foundry EBL service \cite{Gostimirovic2022}, our automated imaging workflow affords this capability to academic cleanrooms. Such training is possible through a foundry service, but each iteration takes much longer and is expensive. The foundry fabricates and measures the test structures and returns the data. The designer then trains the model and creates a corrected layout, which is sent back to the foundry for fabrication. Keeping fabrication and metrology within one facility shortens this loop. We also propose that our fabrication prediction can extend beyond the lithography distortions of a fixed foundry process, to include more complicated process steps such as photoresist reflow.

\begin{figure}[h!]
    \centering
    \includegraphics[width=\linewidth]{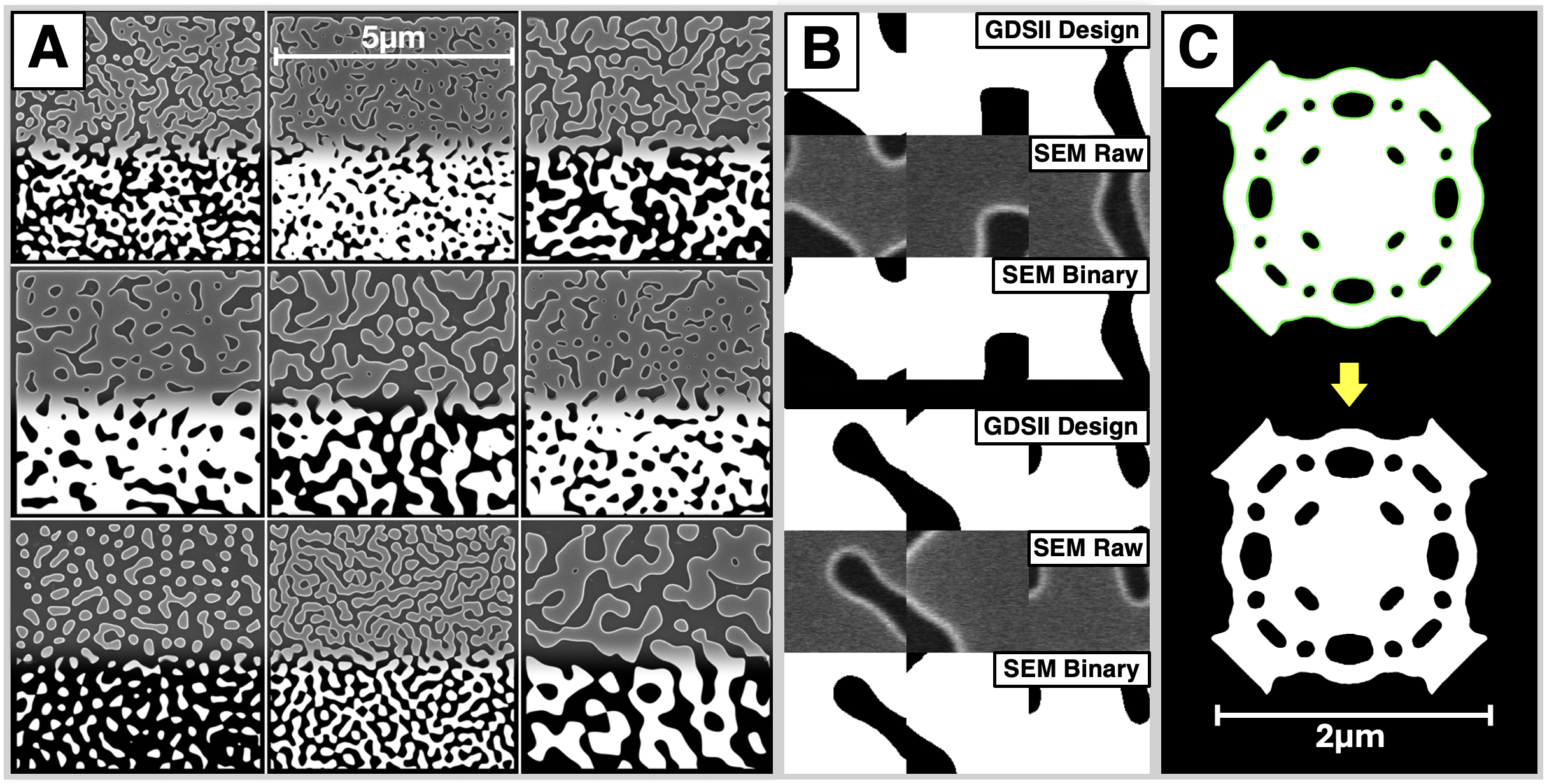}
    \caption{(A) Nine example test patterns in ZEP520A, generated from low-pass and band-pass filtering of noise within $5\times5\,\mu$m$^2$ footprints (scale bar 5\,$\mu$m). Each is shown as an SEM image (top) blended into the binary pattern (bottom). (B) A small set of training patches, broken into three sets of four patches. Each set shows the four original GDSII design binaries, four raw SEM images, and four contour binaries of the raw SEM images. (C) Convolutional neural network prediction of reflow-induced CD broadening (top, green) for a 2\,$\mu$m inverse design photonic crossing, and the `pre-corrector' design change (bottom) suggested so that the resulting reflow can distort the crossing into the original, intended design (scale bar 2\,$\mu$m). Reflow is completed at a set bake temperature of 140\,$^\circ$C.}
    \label{fig:reflowCNN}
\end{figure}

To explore the effect of reflow-induced structural variability, the photoresist glass transition temperature (T$_g$) of ZEP520A by Zeon~Corp. is determined by thin film ellipsometry and compared with the manufacturer's stated value of 105\,$^\circ$C \cite{Zeon2003}. The same technical report gives a post-development bake of 100--140\,$^\circ$C, and we find negligible reflow at 105\,$^\circ$C. From these measurements, we calibrated a reflow recipes at 140\,$^\circ$C that reduces line edge roughness in nanofabricated devices, at the expense of up to $\sim$$10$\,nm of structural variation. This is in line with the use of thermal reflow to reduce edge roughness in the fabrication of waveguides and microring resonators \cite{Sidorov2023}. While models of thermal reflow have been developed \cite{Chia2019,Kim2024}, we find no algorithmic pre-correction of reflow-induced structural variation for arbitrary EBL layouts, comparable to the correction of proximity effect distortion by PEC. We believe this is due to chemistry, thickness, dose, temperature and reflow time all individually affecting the reflowed geometry \cite{Sidorov2023,Kim2024}. Rather than attempt to model the effect physically, we pattern 2700 test structures within $5\times5\,\mu$m$^2$ footprints across 30 different dose/current combinations on a single silicon chip, and reflow the patterns for two different time intervals. This is completed for 60\,s and 90\,s reflow durations at the set bake temperature of 140\,$^\circ$C. The patterns are generated from low-pass and band-pass filtering of noise in line with the methods described in Gostimirovic~et~al.~\cite{Gostimirovic2022}, and can be seen in Figure~\ref{fig:reflowCNN}A. Within a specific dose/current combination, there are 90 different test structures that yield islands whose areas and perimeters vary considerably in order to capture reflow effects across a wide variety of possible features.

When generating these test patterns, scanframes are included in the original GDSII design file, which can be passed directly into \textsc{AutoSEM}. For each reflow duration, six different doses (250--300\,$\mu$C/cm$^2$ in steps of 10\,$\mu$C/cm$^2$) at a 1\,nA beam current with a 300\,$\mu$m aperture are explored with 90 test patterns each, yielding 540 different test images per reflow duration and 1080 in total. Each pattern is imaged in an 8\,$\mu$m scanframe at 2\,nm/px under the conditions given in Section~\ref{subsubsec:methods/image-acquisition}. Unattended image acquisition takes $\sim$$7.5$ hours, and automated post-processing prepares the images for training. With preset scan conditions, reflow structural variation of up to $\sim$$10$\,nm can be reliably observed between design and patterned structures. For convolutional neural network preparation, the images are sliced into overlapping 128$\times$128\,px patches for binarisation and training (a small sample can be seen in Figure~\ref{fig:reflowCNN}B). The result of this training can be seen in Figure~\ref{fig:reflowCNN}C, predicting CD broadening of a 2\,$\mu$m inverse design photonic crossing (top). These broadening results (top, green) can then be used to apply a `pre-corrector' model \cite{Gostimirovic2023,Azimi2024}, suggesting a pre-fabrication design change (bottom) so that the resulting reflow can distort the crossing into the original, intended design.

Without an automated DBM workflow, building up a large and reliable dataset for network training is untenable for an academic nanofabrication facility. When 30 different patterning parameter combinations are varied across a total of 2700 test structures, manually imaging these structures would take an experienced user weeks of continuous imaging to acquire the dataset. Exploring the multiple temperatures across multiple different pattern datasets would require months of continuous imaging.


\section{Discussion}\label{sec:discussion}

\subsection{Closing the fabrication feedback loop}

Automated and comprehensive sweeping of patterning conditions allows the fabrication feedback loop to be closed efficiently and supports best practice process calibration. The automated DBM workflow also makes this accessible to all users within an academic cleanroom facility, regardless of their experience in process calibration. By introducing scanframe locations at the design stage, users can structure comprehensive metrology into their nanofabrication. Automated, unattended image acquisition encourages users to not just acquire images at select on-chip devices, but fully image all relevant design areas. Once imaging is no longer the bottleneck, large datasets of test structures can be designed and fabricated to investigate complex process distortions such as photoresist reflow.

\subsection{Limitations of methodology \& open challenges}

The current limitations in the methodology are the requirements for users to conduct the three-point alignment to a high standard for correct rotation/working distance correction. Inconsistency could lead to systematic error throughout the acquired dataset. Laser height sensing, offered as an option on current Raith pattern generators \cite{RaithMultifunctionalEBL} but not supplied on the base eLine~Plus model that this methodology is based around, would remove part of this limitation. It corrects working distance, focus and sample tilt \cite{RaithAutoRun}, but not in-plane rotation or translation, which still require three-point alignment to the sample markers. A future adaptation of this methodology being explored is to make use of the automatic alignment marker detection in Raith's \textsc{NanoSuite} \cite{RaithMultifunctionalEBL}. Structuring automated alignment marker detection scripts into the existing position list workflow would automate the three-point alignment of chips/wafers. Combining alignment marker detection scripts with laser height sensing would ensure, across any sample analysed by the eLine~Plus, that the same rotation/working distance correction is applied consistently to every dataset acquisition. In this way, academic cleanrooms adopting this automated workflow could choose to take a more custodial role in the process. These facilities could maintain not just the scanframes and alignment-corrected writefields, but the sizes and positions of automated metrology markers that a user must have within their design floorplan to order full-chip/wafer metrology. This would significantly increase automation, shifting the bottleneck from training users in the entire imaging process, to simply training them in loading/unloading procedures.

\subsection{Generalisable workflow \& outlook}

We believe, with the coordination of Raith, that this workflow may be implemented on any pattern generator with the Raith \textsc{NanoSuite}. With minor alterations, it could also run on any SEM with rotation and working distance correction and application programming interface (API) support to scan a position list. The Raith eLine~Plus is just a single, superseded pattern generator to which this workflow has been applied.

Finally, we expect that this workflow may be generalised beyond the silicon photonics use cases described within this methodology, and applied to almost any nanofabrication. Within our facility, users are using the workflow to examine critical dimensions of Josephson junctions. Variation in junction area, and in the nanometre-scale linewidths that define it, shifts the frequency of superconducting qubits, of which Josephson junctions are a constituent non-linear element \cite{Kreikebaum2020,Muthusubramanian2024}. Automatic SEM linewidth metrology over thousands of junctions has already been used to relate junction linewidth to resistance \cite{Pishchimova2023}, and our workflow could be used to refine this metrology and the metrology of many other nanofabrication applications.


\section{Conclusion}

We present an automated DBM workflow that converts a superseded Raith eLine~Plus pattern generator into an unattended CD-SEM, using \textsc{AutoSEM} to automate position list structuring from a GDSII design floorplan and GenISys software to extract key fabrication metrics. Use Case I demonstrates how 400 images of isofocal test structures can be acquired unattended across $\sim$$5.5$ hours and passed through \textsc{ProSEM} and \textsc{Tracer} to estimate the isofocal dose, process blur and process bias of two EBL photoresists. Use Case II highlights how metrology of all 1104 structures on a silicon photonic chip can be organised at the design stage. Use Case III demonstrates that 540 different images per reflow duration, acquired unattended from an array of 2700 test patterns, can be used to train a convolutional neural network that predicts reflow-induced structural variation. The main limitation is the requirement for users to conduct the three-point alignment to a high standard, and structuring Raith's automatic alignment marker detection scripts into the position list workflow is being explored to automate this step. We foresee a future where facilities adopt this automation workflow, adapted for their individual systems, to help users unlock best practice metrology for their nanofabrication using existing facility hardware.

\section*{Acknowledgements}

We acknowledge the traditional owners of the land on which the University of Queensland is situated, the Turrbal and Jagera people. We acknowledge the support of Microscopy Australia at the Centre for Microscopy and Microanalysis at The University of Queensland, and that this work used the Queensland node of the NCRIS-enabled Australian National Fabrication Facility (ANFF). Finally, the authors wish to dedicate this paper to the late Roger Wepf, whose passion for microscopy was the driving force behind this research.

WG, HC, DP, FG, MR, AGW were involved in drafting the paper. WG, HC, ZD were involved in data collection. WG, HC, DP, FG, LH, ZD were involved in data analysis. WG, HC created \textsc{AutoSEM}, LH created \textsc{ProSEM}/\textsc{Tracer} batch analysis and visualisation scripts, KM supplied the JavaScript eLine~Plus automation backend scripts. NT supplied \textsc{Beamer}, \textsc{ProSEM} and \textsc{Tracer} processes. Workflow development was initiated by RW, HC, ZD. Anthropic’s AI model Opus~5.5 was used in this research while coding \textsc{AutoSEM}, as well as OpenAI’s model Codex for \textsc{ProSEM} automation scripting and for batch analysis. Opus~5.5 was used for data visualisation, proof-reading and re-wording during manuscript drafting.

\section*{Funding}

WG is the recipient of a Queensland Government Quantum \& Advanced Technologies Scholarship and funded by the Queensland Government Department of Environment, Tourism, Science and Innovation (DETSI). AGW is the recipient of an Australian Research Council Laureate Fellowship (FL210100045) funded by the Australian Government. This research was partially funded by the Australian Government through the Australian Research Council Centre of Excellence for Engineered Quantum Systems (EQUS, CE170100009).

\section*{Conflict of interest}

We acknowledge the commercial interests of KM, affiliated with Raith Asia Ltd, and NT, affiliated with GenISys GmbH, whose hardware and software (eLine~Plus, EBPG5150, \textsc{NanoSuite}, \textsc{ProSEM}, \textsc{Tracer}, \textsc{Beamer}) are used or described in the article, and whose vendor material is cited. All other authors state no conflict of interest.

\bibliographystyle{unsrtnat}
\bibliography{arxiv_bib}

@article{Orji2018,
  author    = {Orji, N. G. and others},
  title     = {{Metrology for the next generation of semiconductor devices}},
  journal   = {Nature Electronics},
  volume    = {1},
  number    = {10},
  pages     = {532--547},
  publisher = {Springer Science and Business Media LLC},
  year      = {2018},
  doi       = {10.1038/s41928-018-0150-9}
}

@techreport{IRDS2023,
  author       = {{IEEE International Roadmap for Devices and Systems}},
  title        = {{Metrology}},
  institution  = {Institute of Electrical and Electronics Engineers},
  type         = {{IRDS} 2023 Update},
  year         = {2023},
  doi          = {10.60627/FF6X-D213},
  note         = {\doi{10.60627/FF6X-D213}},
  url          = {https://irds.ieee.org/images/files/pdf/2023/2023IRDS_MET.pdf}
}

@misc{GenISysProSEM,
  author       = {{GenISys GmbH}},
  title        = {{ProSEM: Automated Pattern \& Process Metrology}},
  year         = {2026},
  howpublished = {\url{https://prosem.genisys-gmbh.com/features.html}}
}

@misc{GenISysTRACER,
  author       = {{GenISys GmbH}},
  title        = {{TRACER: Electron scattering and process effects quantified}},
  year         = {2026},
  howpublished = {\url{https://www.genisys-gmbh.com/tracer.html}}
}

@misc{GenISysBEAMER,
  author       = {{GenISys GmbH}},
  title        = {{BEAMER: Electron- and Laser-Beam Lithography Software}},
  year         = {2026},
  howpublished = {\url{https://www.genisys-gmbh.com/beamer.html}}
}

@article{Cho2024,
  author    = {Cho, Yunje and others},
  title     = {{Automatic beam optimization method for scanning electron microscopy based on electron beam Kernel estimation}},
  journal   = {Communications Engineering},
  volume    = {3},
  number    = {1},
  pages     = {82},
  publisher = {Springer Science and Business Media LLC},
  year      = {2024},
  doi       = {10.1038/s44172-024-00230-3}
}

@misc{GenISysInSPEC,
  author       = {{GenISys GmbH}},
  title        = {{InSPEC: Integrated SEM Metrology and Inspection}},
  year         = {2024},
  howpublished = {\url{https://www.genisys-gmbh.com/products/inspec.html}}
}

@misc{RaithMultifunctionalEBL,
  author       = {{Raith GmbH}},
  title        = {{PIONEER \textbar{} E-LINE \textbar{} RAITH150: Multifunctional system family for maximum versatility}},
  year         = {2026},
  howpublished = {\url{https://raith.com/products/multifunctional-ebl/}}
}

@misc{RaithProImage,
  author       = {{Raith GmbH}},
  title        = {{ProImage}},
  howpublished = {\url{https://raith.com/technologies/proimage}},
  year         = {2026}
}

@misc{RaithAutoRun,
  author       = {{Raith GmbH}},
  title        = {{AutoRun -- Automated EBL}},
  year         = {2026},
  howpublished = {\url{https://raith.com/technologies/autorun/}}
}

@article{SalvatPujol2019,
  author    = {Salvat-Pujol, Francesc and Villarrubia, John S.},
  title     = {{Conventional vs. model-based measurement of patterned line widths from scanning electron microscopy profiles}},
  journal   = {Ultramicroscopy},
  volume    = {206},
  pages     = {112819},
  publisher = {Elsevier BV},
  year      = {2019},
  doi       = {10.1016/j.ultramic.2019.112819}
}

@article{Lopez2017,
  author    = {Lopez, Gerald G. and others},
  title     = {{Isofocal dose based proximity effect correction tolerance to the effective process blur}},
  journal   = {Journal of Vacuum Science \& Technology B, Nanotechnology and Microelectronics: Materials, Processing, Measurement, and Phenomena},
  volume    = {35},
  number    = {6},
  pages     = {06G505},
  publisher = {American Vacuum Society},
  year      = {2017},
  doi       = {10.1116/1.4995421}
}

@article{Lopez2018,
  author    = {Lopez, Gerald G. and others},
  title     = {{On the trends and application of pattern density dependent isofocal dose of positive resists for 100 keV electron beam lithography}},
  journal   = {Journal of Vacuum Science \& Technology B, Nanotechnology and Microelectronics: Materials, Processing, Measurement, and Phenomena},
  volume    = {36},
  number    = {6},
  pages     = {06JA05},
  publisher = {American Vacuum Society},
  year      = {2018},
  doi       = {10.1116/1.5048206}
}

@article{Gostimirovic2022,
  author    = {Gostimirovic, Dusan and others},
  title     = {{Deep Learning-Based Prediction of Fabrication-Process-Induced Structural Variations in Nanophotonic Devices}},
  journal   = {ACS Photonics},
  volume    = {9},
  number    = {8},
  pages     = {2623--2633},
  publisher = {American Chemical Society (ACS)},
  year      = {2022},
  doi       = {10.1021/acsphotonics.1c01973}
}

@article{Gostimirovic2023,
  author    = {Gostimirovic, Dusan and others},
  title     = {{Improving Fabrication Fidelity of Integrated Nanophotonic Devices Using Deep Learning}},
  journal   = {ACS Photonics},
  volume    = {10},
  number    = {6},
  pages     = {1953--1961},
  publisher = {American Chemical Society (ACS)},
  year      = {2023},
  doi       = {10.1021/acsphotonics.3c00389}
}

@misc{Allresist2026,
  author       = {{Allresist GmbH}},
  title        = {{E-Beam Resist AR-P 6200 series (CSAR 62)}},
  year         = {2026},
  howpublished = {\url{https://www.allresist.com/portfolio-item/e-beam-resist-ar-p-6200-series-csar-62/}}
}

@misc{MicroResist2026,
  author       = {{micro resist technology GmbH}},
  title        = {{ma-N 2400 series -- Negative Tone Photoresist Series}},
  year         = {2026},
  howpublished = {\url{https://microresist.de/en/produkt/ma-n-2400-series/}}
}

@article{Bogaerts2018,
  author    = {Bogaerts, Wim and Chrostowski, Lukas},
  title     = {{Silicon Photonics Circuit Design: Methods, Tools and Challenges}},
  journal   = {Laser \& Photonics Reviews},
  volume    = {12},
  number    = {4},
  pages     = {1700237},
  publisher = {Wiley},
  year      = {2018},
  doi       = {10.1002/lpor.201700237}
}

@article{Ma2024,
  author    = {Ma, Hansi and others},
  title     = {{Inverse-designed 6 $\times$ 6 waveguide crossing for a Benes-based network}},
  journal   = {Optics Express},
  volume    = {32},
  number    = {24},
  pages     = {42908},
  publisher = {Optica Publishing Group},
  year      = {2024},
  doi       = {10.1364/OE.539990}
}

@misc{Zeon2003,
  author       = {{ZEON Corporation}},
  title        = {{ZEP520A Technical Report, Ver. 1.01}},
  year         = {2003},
  howpublished = {\url{https://www.nanophys.kth.se/nanolab/resists/zep520a-7-2.pdf}}
}

@article{Sidorov2023,
  author    = {Sidorov, Fedor and Rogozhin, Alexander},
  title     = {{Thermal Reflow Simulation for PMMA Structures with Nonuniform Viscosity Profile}},
  journal   = {Polymers},
  volume    = {15},
  number    = {18},
  pages     = {3731},
  publisher = {MDPI AG},
  year      = {2023},
  doi       = {10.3390/polym15183731}
}

@article{Chia2019,
  author    = {Chia, Charmaine and others},
  title     = {{Neural network-based model of photoresist reflow}},
  journal   = {Journal of Vacuum Science \& Technology B, Nanotechnology and Microelectronics: Materials, Processing, Measurement, and Phenomena},
  volume    = {37},
  number    = {6},
  pages     = {061604},
  publisher = {American Vacuum Society},
  year      = {2019},
  doi       = {10.1116/1.5116857}
}

@article{Kim2024,
  author    = {Kim, Sang-Kon},
  title     = {{Contact Hole Shrinkage: Simulation Study of Resist Flow Process and Its Application to Block Copolymers}},
  journal   = {Micromachines},
  volume    = {15},
  number    = {9},
  pages     = {1151},
  publisher = {MDPI AG},
  year      = {2024},
  doi       = {10.3390/mi15091151}
}

@inproceedings{Azimi2024,
  author    = {Azimi, Rambod and others},
  title     = {{SEMU-Net: A Segmentation-Based Corrector for Fabrication Process Variations of Nanophotonics with Microscopic Images}},
  booktitle = {2025 IEEE/CVF Winter Conference on Applications of Computer Vision (WACV)},
  pages     = {1528--1536},
  publisher = {IEEE},
  year      = {2025},
  doi       = {10.1109/WACV61041.2025.00156}
}

@article{Kreikebaum2020,
  author    = {Kreikebaum, J. M. and others},
  title     = {{Improving wafer-scale Josephson junction resistance variation in superconducting quantum coherent circuits}},
  journal   = {Superconductor Science and Technology},
  volume    = {33},
  number    = {6},
  pages     = {06LT02},
  publisher = {IOP Publishing},
  year      = {2020},
  doi       = {10.1088/1361-6668/ab8617}
}

@article{Muthusubramanian2024,
  author    = {Muthusubramanian, Nandini and others},
  title     = {{Wafer-scale uniformity of Dolan-bridge and bridgeless Manhattan-style Josephson junctions for superconducting quantum processors}},
  journal   = {Quantum Science and Technology},
  volume    = {9},
  number    = {2},
  pages     = {025006},
  publisher = {IOP Publishing},
  year      = {2024},
  doi       = {10.1088/2058-9565/ad199c}
}

@article{Pishchimova2023,
  author    = {Pishchimova, Anastasiya A. and others},
  title     = {{Improving Josephson junction reproducibility for superconducting quantum circuits: junction area fluctuation}},
  journal   = {Scientific Reports},
  volume    = {13},
  number    = {1},
  pages     = {6772},
  publisher = {Springer Science and Business Media LLC},
  year      = {2023},
  doi       = {10.1038/s41598-023-34051-9}
}

\end{document}